\documentclass[conference]{IEEEtran}
\IEEEoverridecommandlockouts
\usepackage{cite}
\usepackage{amsmath,amssymb,amsfonts}
\usepackage{algorithmic}
\usepackage{graphicx}
\usepackage{textcomp}
\def\BibTeX{{\rm B\kern-.05em{\sc i\kern-.025em b}\kern-.08em
    T\kern-.1667em\lower.7ex\hbox{E}\kern-.125emX}}

\usepackage{comment} %

\excludecomment{IEEEOnly}
\includecomment{arXivOnly}

\usepackage[hyphens]{url}                %
\usepackage[]{hyperref}         %
\usepackage{breakurl}           %
\hypersetup{                    %
  colorlinks,
  linkcolor={green!80!black},
  citecolor={red!70!black},
  urlcolor={blue!70!black}
}
\usepackage[boxed,noend]{algorithm2e} %
\usepackage{amsmath} %
\usepackage{amsthm}
\usepackage{circledsteps} %
\usepackage{cprotect} %
\usepackage{cleveref} %
\usepackage{dblfloatfix} %
\usepackage{float} %
\begin{IEEEOnly}
\usepackage{flushend} %
\end{IEEEOnly}
\usepackage{hhline} %
\usepackage{listings} %
\usepackage{multirow} %
\usepackage{subcaption} %
\usepackage{tablefootnote} %
\usepackage{tabularx} %
\usepackage[euler]{textgreek} %
\usepackage{xcolor} %
\usepackage{tikz} %
\usetikzlibrary{shapes.geometric, arrows, calc, positioning} %

\SetAlCapSkip{1ex}
\SetAlCapSty{AlgorithmCaptionStyle}
\SetAlCapNameSty{AlgorithmCaptionStyle}

\newtheoremstyle{exampstyle}
{0.5\itemsep}
{\itemsep}
{\itshape}
{}
{\bfseries}
{}
{0.5em}
{}

\theoremstyle{exampstyle}
\newtheorem{invariant}{Requirement}
\crefname{invariant}{requirement}{requirements}
\crefname{lstlisting}{listing}{listings}
\crefname{figure}{\figurename}{\figurename}
\Crefname{figure}{\figurename}{\figurename}

\def\JZ#1{}

\newcommand{\ThisSystem}{\mbox{DeTRAP}}

\newcommand{\uRAI}{\mbox{{\textmu}RAI}}
\begin{document}

\begin{arXivOnly}
\bstctlcite{bstctl:nodash}
\end{arXivOnly}

\definecolor{mygray}{rgb}{0.5,0.5,0.5}

\newcommand{\ASMRegister}[1]{\textcolor{red!50!black}{\texttt{#1}}}

\newcommand{\ASMIdentifier}[1]{\textcolor{teal}{\texttt{#1}}}

\newcommand{\ASMInstruction}[1]{\textcolor{blue!80!black}{\texttt{#1}}}

\definecolor{mauve}{rgb}{0.58,0,0.82}

\lstset{
  literate={ö}{{\"o}}1
           {ä}{{\"a}}1
           {ü}{{\"u}}1,
  commentstyle=\color{green!50!black},  %
  keywordstyle=[1]\color{blue!80!black},        %
  keywordstyle=[2]\color{orange!80!black},      %
  keywordstyle=[3]\color{red!50!black},         %
  stringstyle=\color{mauve},                    %
  identifierstyle=\color{teal},                 %
  backgroundcolor=\color{white},   %
  basicstyle=\small\ttfamily,        %
  breakatwhitespace=false,         %
  breaklines=true,                 %
  captionpos=t,                    %
  extendedchars=true,              %
  frame=single,	                   %
  keepspaces=true,                 %
  language=[RISC-V]Assembler,                 %
  numbers=left,                    %
  numberstyle=\scriptsize\color{mygray}, %
  numbersep=1em,
  rulecolor=\color{black},         %
  showspaces=false,                %
  showstringspaces=false,          %
  showtabs=false,                  %
  tabsize=2,	                   %
  xleftmargin=1.4em,              %
  framexleftmargin=1.1em,         %
  framexrightmargin=-0.3em,       %
  escapeinside={/*}{*/}
}

\title{{\ThisSystem}: RISC-V Return Address Protection With Debug Triggers}

\author{\IEEEauthorblockN{Isaac Richter}
\IEEEauthorblockA{\textit{Department of Electrical and Computer Engineering} \\
\textit{University of Rochester}\\
Rochester, New York, USA \\
isaac.richter@rochester.edu}
\and
\IEEEauthorblockN{Jie Zhou\textsuperscript{\hyperlink{JZHOUAffil}{\textdagger}}} %
\IEEEauthorblockA{\textit{Department of Computer Science} \\
\textit{George Washington University}\\
Washington, DC, USA \\
jie.zhou@gwu.edu}
\and
\IEEEauthorblockN{John Criswell}
\IEEEauthorblockA{\textit{Department of Computer Science} \\
\textit{University of Rochester}\\
Rochester, New York, USA \\
criswell@cs.rochester.edu}
}

\maketitle
\begingroup\renewcommand\thefootnote{\textdagger}
\footnotetext{\hypertarget{JZHOUAffil}{Contributions made while at the University of Rochester}}
\endgroup

\begin{abstract}
  Modern microcontroller software is often written in C/C++ and
  suffers from control-flow hijacking vulnerabilities.
  Previous mitigations suffer from high performance and
  memory overheads and require either the presence of
  memory protection hardware or sophisticated program analysis in the
  compiler.

  This paper presents \emph{\ThisSystem} (Debug Trigger Return Address
  Protection).
  {\ThisSystem} utilizes a full implementation of the RISC-V debug
  hardware specification to provide a write-protected shadow stack for
  return addresses.
  Unlike previous work, {\ThisSystem} requires no memory protection
  hardware and only minor changes to the compiler toolchain.
  
  We tested {\ThisSystem} on an FPGA running a 32-bit RISC-V
  microcontroller core and found average
  execution time overheads to be between 0.5\% and 1.9\% on
  evaluated
  benchmark suites with code size overheads averaging 7.9\% or less.

\end{abstract}

\section{Introduction}
\label{sec:introduction}

Modern microcontroller software is mainly written in C and C++.
Unfortunately, these languages are type-unsafe, and programs written therein may
have memory safety vulnerabilities that can be exploited by
control-flow hijacking attacks~\cite{ROP:TOISS12}.
While restricting control flow by enforcing Control-flow Integrity
(CFI)~\cite{CFI:TISSEC09} can mitigate these attacks, previous work has shown
that advanced control-flow hijacking attacks are still possible if the
integrity of function return addresses is not
protected~\cite{OutOfControl:Oakland14,ROPDanger:UsenixSec14,StitchGadget:UsenixSec14}.
Worse, programs on all mainstream architectures, such as x86, ARM, and
RISC-V, all suffer from this problem~\cite{RiscyROP:RAID2022}.

Previous work has taken one of two approaches to mitigate these
sophisticated attacks.  The first approach protects the integrity of
function return
addresses~\cite{Silhouette:UsenixSec20,RECFISH:ecrts19,SSSoK:Oakland19}.
However, all of these systems induce execution time and memory overheads.
Some~\cite{RECFISH:ecrts19} utilize hardware memory protection features that are
intended for privilege domain separation, such as switching to supervisor mode
when manipulating the shadow stack, but otherwise executing in user mode,
requiring a context switch on every call to a non-leaf function.
The most efficient of these systems is
Silhouette~\cite{Silhouette:UsenixSec20} which imposes
1.3\% performance overhead on CoreMark-Pro~\cite{CoreMarkPro:EEMBC} and
3.4\% performance overhead on BEEBS~\cite{BEEBS:arXiv}.  Worse, the code size
overheads for these same benchmarks are 16.5\% and 5.3\%,
respectively, when the size of untransformed code (namely \verb|libc| and
\verb|libm|) are removed from the baseline.

A second approach is to detect corruption of the return address; such
detection must account for the path that control flow has taken
through the call graph in order to mitigate sophisticated control-flow
hijacking attacks.  
\uRAI~\cite{uRAI:NDSS20} is such a system and has comparable performance
overhead to Silhouette, averaging 2.6\%\footnote{{\uRAI}'s overhead is
2.6\% when a compiler transformation that serendipitously improves
performance is also applied to the baseline against which {\uRAI} is
compared. },
but has code size overheads of 54.1\%.
{\uRAI} also requires computing a complete call graph at compile time,
which requires sophisticated whole-program analysis~\cite{DSA:PLDI07}.
These overheads hinder adoption in microcontrollers.
Additional overheads can force manufacturers to choose between
higher security at the cost of utilizing more expensive hardware with
faster processors and more memory or greater cost-efficiency at the
cost of security.

Additionally, we seek a solution that requires no new hardware features.
New hardware features must be thoroughly tested by manufacturers and
ratified by standards bodies before they are implemented and deployed.
Since new hardware support requires ``buy-in'' from multiple entities, a
solution that uses features already
approved by manufacturers and standards bodies is more likely to gain
adoption.

Modern processors, such as ARM~\cite{ARMv7M,ARMv8M} and
\mbox{RISC-V}~\cite{RISCVDebug}, provide sophisticated processor watchpoint
features that can generate a debug watchpoint trap when certain
conditions, configured by software, occur.  Unlike earlier processors,
these new debugging facilities can generate a watchpoint trap
when the program counter or a load or store address is within an arbitrary
range,
or
when the processor is executing a particular instruction.
Furthermore, conditions can be chained together so that a trap occurs
only when multiple conditions are met.
While previous work~\cite{PicoXOM:SecDev20} has employed these
hardware features to implement execute-only memory, we observe that we
can use these features to implement more dynamic security policies,
such as write-protected shadow stacks, which must distinguish
stores that save return addresses from other stores within a program.

In this paper, we leverage these debugging features to build
\emph{Debug Trigger Return Address Protection ({\ThisSystem})}:
a system that combines a novel compilation strategy with modern processor
debug facilities to provide an efficient write-protected shadow stack.
Unlike prior work~\cite{Silhouette:UsenixSec20,PHMon:UsenixSec20},
{\ThisSystem} has minimal
hardware requirements: it requires no memory protection,
address translation, or privilege mode hardware.  Furthermore, {\ThisSystem}
only uses functionality \emph{already specified} in the RISC-V
ISA~\cite{RISCVUnpriv,RISCVPriv,RISCVDebug}; no new hardware needs to pass
through standards committees.
Finally, {\ThisSystem} does not need sophisticated whole-program call graph
analysis.

We prototyped {\ThisSystem} by enhancing
the \mbox{RISC-V} Rocket Core~\cite{RocketChip} to
fully implement the complete RISC-V debugging facilities~\cite{RISCVDebug}
(adding just 0.87\% to the core pipeline)
and by enhancing the LLVM compiler~\cite{LLVM:CGO04} to implement a
write-protected shadow stack using these debugging features.
Our experimental results show that {\ThisSystem} outperforms previous
work such as Silhouette~\cite{Silhouette:UsenixSec20} and
\uRAI~\cite{uRAI:NDSS20}: {\ThisSystem}
incurs execution time overhead of just 0.5\% averaged across
the benchmarks in CoreMark-Pro and 0.8\% for the BEEBS benchmarks evaluated by
Silhouette~\cite{Silhouette:UsenixSec20}, an improvement of 0.5\% and 8.5\%
respectively.  Our results also show that
{\ThisSystem} incurs a code size overhead of 7.9\% and 6.7\%, respectively.
On CoreMark, {\ThisSystem}'s execution time overhead is 1.9\%, an improvement
of 5.7\% against \uRAI~\cite{uRAI:NDSS20}; and it has a flash size
\textit{decrease} of 2.7\%, a $\mathrm{\sim}$40\% improvement.
We further evaluated against Embench~\cite{embench:riscv2019} and found
performance overhead of 1.4\% with a code size \textit{decrease} of 3.5\%.

We evaluated a modification to the rocket core pipeline to implement the parts
of the RISC-V ISA~\cite{RISCVDebug} needed for {\ThisSystem}, and found that
it can be done using just 0.14\% additional cell area (pre-routing) in core
pipeline.

To summarize, the main contributions of this paper are:

\begin{itemize}
  \item
  The design of the first system that uses modern processor debugging
  facilities to implement efficient write-protected shadow stacks

  \item
  A return address integrity system that can support unmodified
  untrusted leaf functions in precompiled code and handwritten assembly

  \item
  A {\ThisSystem} prototype that implements our design on the
  \mbox{RISC-V} Rocket Core~\cite{RocketChip} 

  \item
  An evaluation of the hardware changes to Rocket Core that would be needed
  to support {\ThisSystem}.

  \item
  An evaluation of {\ThisSystem}'s performance and code size
  overheads, showing that {\ThisSystem} provides the same protection
  as previous work with less performance and memory overhead.  Unlike
  previous work~\cite{Silhouette:UsenixSec20,uRAI:NDSS20}, our
  evaluation methodology removes serendipitous code layout changes
  as the source for improved performance in our evaluation results.
\end{itemize}

\section{Background on \mbox{RISC-V} Debug Triggers}
\label{sec:background}

\label{sec:background-triggers}

The \mbox{RISC-V} architecture~\cite{RISCVDebug} provides a rich set of
primitives for specifying the
conditions under which the processor should trigger a breakpoint exception.
Breakpoints can be configured to fire prior to entering a trap handler,
after a configurable number of instructions has been executed, or
based on a comparison against
a program counter,
load/store address,
instruction opcode,
and/or
data value loaded from or stored to memory.

Breakpoint comparisons are not limited to equality checking;
comparisons can also be configured to trigger if a value is less
than, greater than or equal to, or unequal to another
value~\cite{RISCVDebug}.
It is also possible to define a \textbf{bitmask match}, where selected
mask bits of an input value are checked against the same bits of a
stored pattern.
This functionality can be used to raise an exception if a specific
instruction is to be executed regardless of the registers encoded in the opcode.

What makes \mbox{RISC-V} breakpoints particularly powerful is that multiple
debug triggers
can be chained together so that the processor only traps if all conditions in
the chain are met~\cite{RISCVDebug}.
This feature permits trapping on conditions that are more complex than can be
described by a single comparison.
For example, trapping when executing code within an arbitrary region can be done
by using two triggers,
one each for the region's lower and upper bounds.
The Debug ISA also allows chained triggers to mix and match what is being
compared.
For example, a data value trigger can be chained with a store address trigger,
trapping when the code attempts to write a specific value to a specific
memory location.

Together, these features provide efficient conditional breakpoints,
alleviating the need to check for conditions in the
exception handler.
Furthermore, because debug triggers operate in parallel with execution,
they perform checks without any per-check performance penalty.

To balance functionality with performance and cost,
current implementations generally only include a few triggers:
SiFive's FU540~\cite{FU540:SiFive} and FU740~\cite{FU740:SiFive} chips
only include two debug triggers per hardware thread (hart).
Since triggers are configured per-hart,
this substantially limits their usability, as
applying a policy across all harts requires duplicating the
configuration across them as well, so all policies targeting these devices
must collectively
fit into just two triggers.

Moreover, implementations are not required to include all functionality from
the specification.  For example, SiFive's chips only support matches against the
program counter or load/store address and do not support bitmask
matches~\cite{FE310:SiFive,FU540:SiFive,FU740:SiFive}; we know of no implementation that matches
against the instruction opcode or loaded/stored data value.
Furthermore, due to hardware tradeoffs, the debug
specification~\cite{RISCVDebug} anticipates that implementors may want to
restrict the complexity of supported triggers.
Indeed, many
implementations~\cite{RocketChip,FE310:SiFive,FU540:SiFive,FU740:SiFive} limit
 chaining to just two registers.
Supporting longer chains  requires
additional chip
area and could reduce the maximum pipeline clock frequency.

As \Cref{sec:design-triggers} discusses,
{\ThisSystem}'s design is intended for single-core microcontrollers, similar to
SiFive's FE310, which supports 8 triggers on its single
hart~\cite{FE310:SiFive} and allows up to four two-trigger-chain rules to be
defined.

\section{Threat Model}
\label{sec:threat}

Our system protects a single embedded bare-metal application running without
an operating system kernel or supervisor.
For simplicity, we assume that the application is single-threaded and does
\begin{IEEEOnly}
not utilize traps to modify control flow or to context switch to
\end{IEEEOnly}
\begin{arXivOnly}
not utilize traps to modify control flow (e.g., as Linux signal
handlers can~\cite{UnderstandingLinux2:BovetCesati}) or to context switch to
\end{arXivOnly}
other tasks, threads, or processes.
As with many embedded applications and processors, we assume a single address
space application running in privileged mode without any hardware memory
protection mechanisms.
The single application is benign and has no runtime-loadable code, but may have
exploitable spatial~\cite{Ret2Libc:RAID11,ROP:TOISS12} and
temporal~\cite{AfekSharabani:BlackHat07} memory safety errors that can corrupt
control data such as return addresses and function pointers.
Non-control data attacks~\cite{ShuoChen:USENIX05} are out of scope.

Physical attacks, such as connecting an external debugger or
modifying the data on volatile or non-volatile memories, are also out of scope.
The damage from attacks on non-volatile storage can be mitigated through
signature checking implemented by trusted boot running from non-reprogrammable
mask ROM~\cite{OpenTitanSecurityModelSpec,OpenTitanAnalysis:Nordic2021}.
We do not mitigate attacks via separately-programmed
devices that can autonomously modify system memory, except to the extent that
such requests may be initiated after the processor writes to a memory-mapped
register of the peripheral (e.g., DMA engines~\cite{piegdon2007hacking}).

\section{Design}
\label{sec:design}

{\ThisSystem} provides a write-protected compressed shadow stack
which securely stores return addresses.  Such a system has
\textbf{trusted code}, which is permitted to save return addresses to
the shadow stack, perform I/O operations, and write to
security-critical data needed for {\ThisSystem}'s operation;
the rest of the code is \textbf{untrusted code} that should be unable
to write to the shadow stack or security-sensitive registers.

In this section, we first present the requirements that {\ThisSystem} must
meet to enforce return address integrity (\Cref{sec:design-invariants}).
 We then explain how {\ThisSystem} lays out the address space to minimize the
number of debug triggers it needs (\Cref{sec:design-memlayout})
and how it configures those debug registers to prevent untrusted code
from modifying the shadow stack or other security-critical
data (\Cref{sec:design-triggers}).
Next, we explain how {\ThisSystem} transforms code to implement
a write-protected shadow stack
(\Cref{sec:design-runtime-trap-handler,sec:design-compiler-forward-CFI}).
Finally, we discuss additional precautions that {\ThisSystem} takes
to ensure that its protections will operate as intended
(\Cref{sec:design-SSP}--\Cref{sec:design-scanner}).

\begin{figure*}[t]
    \centering
    \includegraphics[scale=.8]{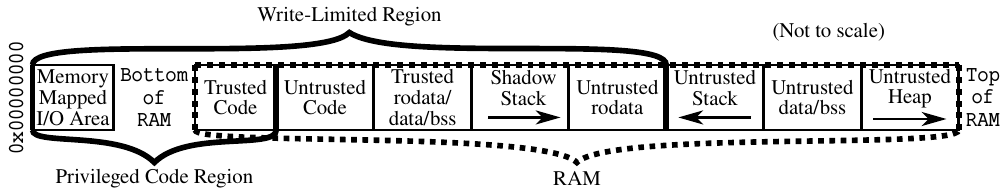} %

  \caption{{\ThisSystem} memory layout for systems without separate
  code and data memories
    }
  \label{fig:memlayout}
\end{figure*}

\subsection{Security Requirements}
\label{sec:design-invariants}

Our design employs a compressed shadow stack~\cite{RAD:ICDCS2001}.
Silhouette~\cite{Silhouette:UsenixSec20} summarizes three high-level invariants
that any shadow stack-based approach for return address integrity must maintain:
(1) A return address is stored either in the shadow stack or in a register
    that is never spilled to memory.
(2) The shadow stack and the register for return addresses cannot be corrupted.
(3) A function's epilogue always retrieves the
    return address that is stored by the function's prologue.
For {\ThisSystem}, we identify five security requirements that must be met
to maintain the three invariants.

First, return addresses are always stored to a trusted location for use by
function epilogues.
Specifically:

\begin{invariant}
  Return addresses used for control flow can only be stored in a dedicated CPU
  register and the shadow stack, and the writes of return addresses can
  only occur in a function's prologue.
  \textnormal{(\Cref{sec:design-return-address-handling,sec:design-scanner})}
  \label{inv:sswrite}
\end{invariant}

Second, a write-protected shadow stack provides no protection unless
a function's epilogue reads the return address from the correct
location within the shadow stack.
{\ThisSystem} uses a shadow stack pointer (SSP) which reads/writes
return addresses from/to the shadow stack. The SSP is stored
in a dedicated hardware register.
However, this design could inadvertently access an incorrect return address or
an arbitrary value from outside the
write-protected shadow stack if the SSP points to the wrong location.
This leads to

\begin{invariant}
  Return addresses must always be retrieved from an uncorrupted CPU register
  or via an uncorrupted shadow stack pointer.
  \textnormal{(\Cref{sec:design-return-address-handling,sec:design-runtime-trap-handler,sec:design-SSP})}
  \label{inv:ssread}
\end{invariant}

Additionally, as {\ThisSystem} protects the shadow stack using the processor's
debugging facilities, the following requirement must be met:

\begin{invariant}
  Software cannot reconfigure the processor's debug registers to modify
  its trap conditions.
  \textnormal{(\Cref{sec:design-scanner})}
  \label{inv:conf}
\end{invariant}

Furthermore, we need to ensure that forward control-flow transfers (e.g. calls
via function pointers) can only target predetermined destinations.
Not only does this mitigate various control-flow hijacking attacks, e.g.,
return-to-libc attacks~\cite{Ret2Libc:RAID11}, but it also
permits {\ThisSystem} to use code scanning techniques~\cite{PNaCL:UsenixSec10,
NestedKernel:ASPLOS15,Hodor:Usenix19} to ensure that
untrusted code does not use instructions that our design deems unsafe.
Therefore:

\begin{invariant}
  Indirect function calls/jumps always branch to the beginning of a function,
  and intra-function indirect jumps always use a destination address loaded from
  their precomputed jumptable.
  \textnormal{(\Cref{sec:design-compiler-forward-CFI})}
  \label{inv:cfi}
\end{invariant}

Finally, as previous work has noted~\cite{HyperSafe:Oakland10,KCoFI:Oakland14},
restricting control flow is useless if an attacker can modify executable code
or CFI metadata.  {\ThisSystem} therefore enforces:

\begin{invariant} All executable code and any data used to enforce CFI cannot be
  corrupted.
  \textnormal{(\Cref{sec:design-memlayout,sec:design-triggers,sec:design-runtime-trap-handler})}
  \label{inv:rodata}
\end{invariant}

\subsection{Memory Layout}
\label{sec:design-memlayout}

Binary executables typically have separate sections for
code, read-only initialized data (rodata), initialized writable data (data), and
uninitialized writable data (bss).  {\ThisSystem} must prevent corruption of
code and security-critical data.
{\ThisSystem} therefore further divides the data sections to
distinguish between those whose integrity it must protect, i.e.,
sections for which corruption would violate a security requirement,
and those that can be modified safely by untrusted code.

Thus, we have both \textbf{untrusted stack} and \textbf{shadow stack} sections,
as well as the \textbf{untrusted data/bss} and \textbf{trusted data/bss}
sections.  Since the rodata section may include data used for
control flow, such as lookup tables for \texttt{switch} statements,
{\ThisSystem} must also protect its integrity.
Most systems include a \textbf{memory mapped I/O (MMIO)} area for
access to peripherals and configuring the processor.
As some peripherals, such as direct memory access (DMA) engines~\cite{piegdon2007hacking},
could be used to violate invariants, {\ThisSystem} must also protect
the MMIO area from untrusted writes.

Like with data, {\ThisSystem} separates code into two sections:
the \textbf{trusted code} section and the \textbf{untrusted code} section,
respectively.
{\ThisSystem} prevents untrusted code from corrupting the shadow stack,
trusted data/bss, rodata, and both code sections.
Thus, each section has two {\ThisSystem}-specific attributes: (1) whether
the section must be \emph{write-limited} because corrupting it
could violate a requirement, and (2)
if the section contains \emph{privileged code}, comprising 
instructions that are trusted not to violate requirements when writing to memory.

To reduce the number of debug triggers needed, {\ThisSystem} groups sections
with identical protection requirements into contiguous regions.
We thus define contiguous write-limited and privileged code
regions and their implied complements (non-write-limited and
unprivileged code), and lay out the sections within their respective
regions.

\Cref{fig:memlayout} shows {\ThisSystem}'s memory layout.
{\ThisSystem} needs to define a trigger chain that detects whether an
instruction in the \emph{unprivileged code region} attempts to write into the
the \emph{write-limited region}.  Since we target a system that only
allows up to two triggers to be chained together, we can only use a
single trigger to match against each region.
We anchor each region at the bottom or top of the address space
and use $<$ and $\ge$ trigger matches, respectively.

Anchoring the write-limited region at the beginning of the address space
has the added benefit of automatically including the memory mapped I/O area,
protecting against DMA attacks~\cite{piegdon2007hacking}.
Additionally, by placing
the non-write-limited region at the top of the address space and placing
the untrusted stack at the bottom of the non-write-limited region,
{\ThisSystem} can provide
free detection of stack overflow in untrusted code; stack overflows attempt to
write to the write-limited region, generating a trap.

{\ThisSystem} also anchors the privileged code region to the bottom of the
address space, and the privileged code region includes the memory mapped I/O area.
This has the benefit that ROM routines, which are vetted during hardware design,
can directly write to MMIO addresses.

While there are no hardware-enforced memory protections against execution
outside of code regions, {\ThisSystem}'s forward CFI protection
(\Cref{sec:design-compiler-forward-CFI}) makes it impossible to transfer
control flow outside of the code sections.  Hardware memory protection
to control which memory sections are executable is therefore unneeded.

\begin{table}[t]
	\centering
    \footnotesize{
    \caption{Configured Debug Triggers}
    \begin{tabularx}{\columnwidth}{|X|c|c|}
        \hline
        \multicolumn{3}{|c|}{\bfseries Write Protection Chain (Two triggers)} \\
        \hline
        Program Counter & $\ge$ & Bottom of Untrusted Code Section \\
        Write Address & $<$ & Bottom of Untrusted Stack Section \\
        \hhline{===}
        \multicolumn{3}{|c|}{\bfseries Shadow Stack Overflow Prevention Chain
        (One trigger)} \\
        \hline
        Write Address & $=$ & Top of Shadow Stack \\
        \hline
    \end{tabularx}
    \label{table:debugtriggers}}
\end{table}

\subsection{Debug Triggers}
\label{sec:design-triggers}

We observe that we can leverage the debug triggers, as discussed in
\Cref{sec:background-triggers}, to enforce a security policy
as a chain of trigger conditions.
Since software running on the processor can configure the debug hardware,
we can exploit the debugging facilities not for debugging but to enforce
security policies.
As \Cref{table:debugtriggers} shows, we use three debug triggers in two
chains---one chain with two triggers and the other trigger by itself---to
implement policies that protect {\ThisSystem}'s sensitive data.

To properly protect the write-limited region, the processor must
trap when code outside the privileged code region attempts to modify data in
the write-limited region. {\ThisSystem} chains two triggers together to enforce
this policy.
The first trigger matches the program counter greater than the top of the
privileged code region, and
the second trigger matches memory write addresses below the top of the
write-limited region. The overall effect is that
if untrusted code attempts to modify the write-limited region, the
processor will raise a breakpoint exception.
The runtime's trap handler (\Cref{sec:design-runtime-trap-handler})
will detect that exception and take corrective action.

Additionally, as discussed below in \Cref{sec:design-SSP},
we use a debug trigger to detect attempts by any code to write to the topmost location
in the shadow stack, preventing the imminent overflow thereof.

\begin{lstlisting}[
  float=b,
  language={[RISC-V]Assembler},
  captionpos=b,
  caption={{\ThisSystem} Function Prologue},
  label=lst:prologue]
.section .text
foo: /*\label{lst:prologue-label}*/
  j foo$trampoline /*\label{lst:prologue-call}*/
foo$postjump:           #post-jump label/*\label{lst:prologue-postjump}*/
  addi sp, sp, -FRMSIZE #original prologue/*\label{lst:prologue-continue}*/
\end{lstlisting}

\subsection{Return Address Handling}
\label{sec:design-return-address-handling}

\begin{lstlisting}[
  float=b,
  language={[RISC-V]Assembler},
  captionpos=b,
  caption={Return Address Save Trampoline},
  label=lst:trampoline-function]
.section .trusted.text
foo$trampoline:
  sw   ra,  0(x3)   /*\label{lst:trampoline-direct-store}*/
  addi x3, x3, 4    /*\label{lst:trampoline-direct-add}*/
  j    foo$postjump /*\label{lst:trampoline-direct-jump}*/
\end{lstlisting}

{\ThisSystem}'s compiler generates the function prologue and epilogue code
that saves and restores return addresses.
For leaf functions, i.e., functions that do not call other functions,
the prologue saves the return address in an ABI-defined register that
is reserved for return addresses.  Since the compiler generates no
other code that writes to this reserved register, the return address remains
safe.
Non-leaf functions, though, must use special prologue and epilogue code
to safely save the return address to and restore it from the shadow stack.

Since function prologues write return addresses to the shadow stack,
they must utilize code (called a \emph{trampoline}) within the trusted
code region.  For each non-leaf function, the {\ThisSystem} compiler generates
runtime trampolines, located within the trusted code
region, that writes the return address to the shadow stack.
When a function (e.g., \ASMIdentifier{foo}\nolinebreak) needs to save its
return address, its prologue (see \Cref{lst:prologue}) first
calls its associated trampoline
(\ASMIdentifier{foo\$trampoline}\nolinebreak) in the trusted code segment.
This trampoline (see Listing~\ref{lst:trampoline-function}) saves the return
address, which was set by the \ASMInstruction{call} instruction,
to the write-protected shadow stack, and
then jumps back to the start of its associated function's original prologue
(\ASMIdentifier{foo\$postjump}\nolinebreak).

To optimize performance, when calling a function with a trampoline, the call
is modified to directly branch to the trampoline
(e.g., \ASMIdentifier{foo\$trampoline}).
If a function has no external linkage and all calls to it are replaced by calls
to its trampoline code, the compiler can further reduce code size by removing
the jump to the trampoline
(\Cref{lst:prologue} \mbox{lines~\ref{lst:prologue-label}-\ref{lst:prologue-call}})
from the function's prologue.

For non-leaf functions, {\ThisSystem} inserts code into the function epilogue
to restore the
return address from the shadow stack (see \Cref{lst:epilogue}).
Unlike function prologues, epilogues need no trampoline in the trusted
code region because they do not write to the shadow stack.
{\ThisSystem}'s code generator, CFI (\Cref{sec:design-compiler-forward-CFI}),
and code scanner (\Cref{sec:design-scanner}) ensure that the shadow
stack pointer is never corrupted, guaranteeing that the epilogue always
loads the correct return address from the write-protected shadow stack.

\subsection{Trap Handling}
\label{sec:design-runtime-trap-handler}

If unprivileged code attempts to write to the write-limited region or the
shadow stack overflows, the debug triggers described in \Cref{sec:design-triggers}
will cause a trap.
To ensure that this trap is handled properly by trusted code, {\ThisSystem}
therefore performs initial handling of all traps.
If the trap is an exception within trusted code, or was caused by an
attempt to violate {\ThisSystem}'s protections, the handler will
terminate execution.
However, not all traps are caused by violations of {\ThisSystem}'s
security requirements, such as timer interrupts.
In these cases, {\ThisSystem}'s handler will create a trap frame and invoke the
application's (untrusted) handler.

To protect against corruption of potentially sensitive processor state, the trap frame
is always saved to the shadow stack.
When handling interrupts, this ensures that the untrusted handler cannot
potentially subvert sensitive operations in-progress (e.g., return address
handling code within prologues or epilogues).
The untrusted handler can still modify data outside the write-limited region to,
for example, set a flag or copy data into or out of an I/O buffer.

For exceptions in unprivileged code, the trap frame is copied onto the untrusted
stack for the
application's handler to modify, with some limitations.
The shadow stack pointer and return address registers are always restored from
the trap frame on the shadow stack, ensuring that they cannot be corrupted.
The untrusted handler can modify the program counter only to increment it to
the next instruction, for example when
emulating a floating-point instruction on a system without a floating-point
unit.
Any other modification of the PC could violate control-flow, and is prohibited.
If the untrusted handler attempts a prohibited modification, this is treated
like any other violation, and execution is terminated.

\begin{lstlisting}[
  float=b,
  language={[RISC-V]Assembler},
  captionpos=b,
  caption={{\ThisSystem} Function Epilogue},
  label=lst:epilogue]
# Restore original stack pointer /*\label{lst:epilogue-original}*/
addi sp, sp, FRMSIZE
# Load return address from shadow stack
lw   ra,  -4(x3)/*\label{lst:epilogue-load}*/
# Decrement shadow stack pointer
addi x3, x3, -4 /*\label{lst:epilogue-add}*/
jr ra    # Original Function Return/*\label{lst:epilogue-return}*/
\end{lstlisting}

{\ThisSystem} includes a code scanner (see \Cref{sec:design-scanner}), which
verifies that
untrusted code, including the application's trap handler, does not include the
trap return (\ASMInstruction{mret}) instruction.
\JZ{Add citation for mret.}
Because it is not safe for the untrusted handler to modify the shadow stack pointer,
the application cannot use the trap
handler to transfer control between threads, as in an interrupt-driven
scheduler.
If it were necessary to enable context switching, methodology similar
to that of Kage~\cite{Kage:UsenixSec22} could be applied---the
trusted code could save per-task state in a per-task shadow stack within the
write-limited region and switch between them on-demand.

\subsection{Forward-edge Control Flow Integrity}
\label{sec:design-compiler-forward-CFI}
\label{sec:design-compiler}

{\ThisSystem}'s debugger triggers enforce shadow stack integrity
(\Cref{sec:design-triggers}), and {\ThisSystem} ensures that trap handlers
cannot corrupt the SSP (\Cref{sec:design-runtime-trap-handler}).
However, to completely prevent the exploitation or misuse of the SSP,
{\ThisSystem} also must ensure that forward-edge control flow cannot
transfer to the middle of function prologues and epilogues.
Specifically, inter-function forward branches must jump to the first
instruction of a function's prologue,
and intra-function branches must jump to a valid location within the function.
Additionally, on processors supporting variable-length instructions,
such as RISC-V's ``C'' Compressed Instruction Extension~\cite{RISCVUnpriv},
{\ThisSystem} must ensure that branches jump to the beginning of an
intended instruction, as there might be a coincidental and valid sequence
of instructions that is an offset from the intended
instructions that could subvert {\ThisSystem}'s security.

For intra-function control-flow integrity, specifically, indirect jumps
from {\tt switch} statements, LLVM---the compiler upon which {\ThisSystem}
is based---compiles them to use a bounds-checked jumptable.
\JZ{IR, please check if the previous statement is correct.}
For indirect function calls, {\ThisSystem} uses LLVM's
\mbox{\texttt{icall-cfi}}~\cite{CFICompiler:SEC14,CFI:Clang}, which provides
type-based CFI, meeting more than the minimum requirements above
(more details in \Cref{sec:impl-compiler}).
A more fine-grained CFI could provide better protection against forward-edge
threats, such as call-oriented~\cite{sadeghi2018PCOP} and
function-reuse~\cite{guo2018function} attacks,
but is unnecessary for reverse-edge protection.

Forward-CFI can prevent mismatches between function prologues and epilogues
for most programs. However, misuses of {\tt setjmp/longjmp} may disrupt
the balance.  Since {\tt setjmp/longjmp} are infrequently used in programs
for embedded systems, we provide our design for handling them
\begin{arXivOnly}
in \Cref{appendix:setjmp}.
\end{arXivOnly}
\begin{IEEEOnly}
in our technical report~\cite{richter2024DeTRAParXiv}.
\end{IEEEOnly}

\subsection{Shadow Stack Overflow and Underflow}
\label{sec:design-SSP}

{\ThisSystem} uses a shadow stack pointer to read/write return addresses from/to
the shadow stack, and keep it from being corrupted.
Although the {\ThisSystem} code scanner (\Cref{sec:design-scanner}),
ensures that the SSP is only modified in trusted code and function epilogues
(\Cref{sec:design-return-address-handling}),
it is also necessary to ensure that the SSP cannot underflow or overflow.
While it would be possible to add bounds checks for overflow after each
increment, we instead use an additional debug trigger that matches on writes to
the last entry of the shadow stack.
Because writes to the shadow stack are strictly incremental, this is sufficient
to detect an overflow without any runtime penalties.
Underflow would imply either corruption of the shadow stack pointer, which is
checked for by the code scanner, or
a violation of proper control flow (such as illegal execution of function
prologue/epilogue code), which is handled by {\ThisSystem}'s CFI (\Cref
{sec:design-compiler-forward-CFI}).
This also allows {\ThisSystem} to avoid bounds checks for underflow.

\subsection{Code Scanning}
\label{sec:design-scanner}

After compiling and linking a program, {\ThisSystem} runs a code
scanner on the generated executable and warns about any vulnerabilities
the code scanner discovers.  This code scanner provides two
critical services.  First, as all code runs in the processor's
privileged mode, the code scanner ensures that the program does not
use privileged instructions to bypass {\ThisSystem}'s protections.
Second, the code scanner ensures that \emph{all} native code
(code generated by the {\ThisSystem} compiler, assembly code written by hand,
and code generated by other compilers) does not break {\ThisSystem}'s
security guarantees.

\subsubsection{External Code}
External precompiled code and handwritten assembly must either use
{\ThisSystem}'s return address handling
(\Cref{sec:design-return-address-handling}), handwritten or
generated via {\ThisSystem}'s compiler, or consist only of functions
that keep the return address in the \ASMRegister{ra} register (e.g.:
leaf functions).
Otherwise, the code scanner will detect unsafe loading of the return address.
It is the user's responsibility to ensure that any linked external code
that is in the trusted code section does
not violate {\ThisSystem}'s requirements.

\subsubsection{Configuration Protection}
Untrusted code must not modify the debug trigger or trap handler
configurations as doing so could nullify {\ThisSystem}'s protections.
The debug trigger and trap handler configurations are governed on RISC-V by
Control and Status Registers (CSRs)~\cite{RISCVDebug,RISCVPriv} configured
via the \ASMInstruction{CSRR*}
instructions~\cite{RISCVUnpriv} that perform a read-modify-write operation.
The CSR to be modified is encoded as an immediate value embedded in the opcode.
The code scanner assumes that any CSR instruction that is not an atomic bit 
set/clear instruction with a hard-coded zero input will modify its targeted CSR.
If the code scanner finds a \ASMInstruction{CSRR} instruction that
modifies CSRs governing debugging and trap handling, it rejects the program;
all other CSR modifications are permitted.

\subsubsection{Call and Return Verification}
We designed {\ThisSystem} so that \emph{all} native code loaded on to
the system follows {\ThisSystem}'s requirements.  This includes code
compiled by the {\ThisSystem} compiler and \emph{external code} such
as hand-written assembly language code and library code compiled by other
compilers e.g., a C standard library compiled by GCC.  To this end,
the {\ThisSystem} code scanner performs the following checks on
\emph{all} native code linked into the final binary executable.

First, the code scanner verifies that all indirect branches, including those
in assembly and precompiled code, are preceded by the appropriate CFI checks
as \Cref{sec:design-compiler} describes.  Second, the code
scanner ensures that either the \ASMRegister{ra} register has not been
modified or that it has been spilled and reloaded from the shadow
stack as \Cref{sec:design-return-address-handling} describes.
Additionally, the code scanner verifies that only function epilogue
code modifies the shadow stack pointer register and that it does so
only by decrementing the register by the correct amount
(as shown in \Cref{lst:epilogue}).
Third, the code scanner verifies that only trusted code uses the
trap return instruction \ASMIdentifier{mret}~\cite{RISCVPriv}.

There are some functions that do not follow {\ThisSystem}'s
conventions but are still safe to use, e.g., indirect jumps in the {\tt memset()}
function in {\tt libc}.  The code scanner permits a developer who has
vetted such jumps to add them to a whitelist with their destinations.
This allows the scanner to confirm that the functions otherwise meet
{\ThisSystem}'s requirements.

\section{Implementation}
\label{sec:impl}

Our implementation is based on a purpose-built runtime and a modified version
of Clang and LLVM~\cite{LLVM:CGO04} \mbox{15.0.7}.
We also enhanced the Rocket core~\cite{RocketChip} \mbox{RISC-V} processor to
implement a more recent version of the debug trigger ISA~\cite{RISCVDebug}
.
As our benchmarks do not use
it, we did not implement the \texttt{setjmp}\slash
\begin{arXivOnly}
\texttt{longjmp} handling from \Cref{appendix:setjmp}.
\end{arXivOnly}
\begin{IEEEOnly}
\texttt{longjmp} handling.
\end{IEEEOnly}

\subsection{Rocket Core Modifications}
\label{sec:impl-rocket}

The upstream Rocket breakpoint module is based on the 0.13 draft of the
\mbox{RISC-V} Debug Support Specification.
Implementations are free to support as little of the specification as they want,
using write-any-read-legal (WARL) semantics such that a read-back of the
configuration register will reflect only what is supported.
Unfortunately, Rocket's breakpoint module implementation
does not properly support combining both program counter (PC) and memory triggers
into the same chain, even though its WARL read-back implies it should.
This deviation from the specification is undocumented and
prevented {\ThisSystem} from working properly.
Because {\ThisSystem} needs this functionality,
we modified the implementation to properly support such a chain.

In the Rocket core pipeline, %
the breakpoint module effectively has two breakpoint units (BPUs) that eavesdrop
on the outputs from each stage of execution.
The PC BPU monitors the instruction fetch (IF) stage to determine if a PC trigger
should fire; the MEM BPU checks the input of the memory (MEM) stage --- 
the output of the execute (EXE) stage --- for matches
against memory read or write triggers.
For chained triggers, all triggers in the chain must match during the same
cycle for an exception to be raised.
However, each instruction is executing in only a single pipeline stage
at a time.  Consequently, in any given cycle, the breakpoint module is
examining the behavior of \emph{different} instructions in the PC and MEM
BPUs.  What {\ThisSystem} needs is to have the MEM stage generate a trap \emph{if} the
instruction matched a PC trigger when it was examined by the PC BPU a few
cycles earlier.

We fixed this problem
by adding pipeline registers to the outputs of the instruction decode (ID) and
EXE stages to track
whether individual triggers matched the instruction address.
These PC-based ``pretriggers'' then feed back into the breakpoint module
alongside the memory stage inputs and are combined with the memory triggers to
ensure that mixed chains properly raise exceptions.
Our evaluation in \Cref{sec:performance-utilization} shows that this change
uses negligible
additional area and energy and brings the implementation into compliance
with the specification.

\subsection{Shadow Stack Implementation}
\label{sec:impl-ss}

We modified the Clang/\mbox{LLVM} compiler to implement the write-protected
shadow stack described in \Cref{sec:design-return-address-handling}.
Our modification of function prologues and epilogues is based on Clang's
ShadowCallStack~\cite{ShadowCallStack:Clang}.
Since we built {\ThisSystem} before the RISC-V ABI designated \ASMRegister{x3}
as a platform register~\cite{RISCVABIPlatformRegister,RISCVABIX3Usage}
and ShadowCallStack adopted it as the shadow stack
pointer register~\cite{LLVMRISCVSCSX3},
our implementation uses \ASMRegister{x18} as the shadow stack pointer
register like the original \mbox{RISC-V} ShadowCallStack implementation., unlike
the code shown in \Cref{lst:epilogue,lst:trampoline-function}.
Our implementation writes a copy of the return address to both the shadow
stack and the original untrusted stack; returns use the
write-protected copy from the shadow stack.  This implementation allows
existing code that reads the return address from the stack,
such as \texttt{\_\_builtin\_return\_address}, to function without modification.
{\ThisSystem}, also like ShadowCallStack~\cite{ShadowCallStack:Clang},
merely uses the return address on the write-protected shadow stack
on function return and does not check whether
the return address restored from the shadow stack matches the return
address on the original untrusted stack.

\subsection{Forward-Edge CFI Implementation}
\label{sec:impl-compiler}

To ease implementation of forward-edge control flow protection,
we used Clang/LLVM's existing indirect function call
checking \verb|-fsanitize=cfi-icall|~\cite{CFICompiler:SEC14,CFI:Clang}.
At compile time, this CFI creates jumptable entries for each
function that is address-taken, sorted by the function type signature;
when taking the address of a function, it then substitutes the address of the
jumptable entry instead.
Indirect calls are rewritten to verify that the
pointer is aligned with and in-bounds of those jumptable entries that
match the expected type signature, and then the function is called via the jump
table entry.
This form of CFI exceeds the minimum requirements for
forward-edge control flow identified in \Cref{sec:design-compiler-forward-CFI}.
If greater precision on forward-edge control flow is desired, an
alternative {\ThisSystem} implementation can use other forward-edge CFI
mechanisms (e.g., label-based CFI~\cite{CFI:TISSEC09} with a precise call
graph).

\subsection{{\tt nospill} Attribute for CFI-Sensitive Data}
\label{sec:impl-compiler-nospill}

CFI sensitive data is data that is used to check the destinations of
indirect branches, including
constant values used in CFI run-time checks and {\tt switch} statement jump
calculations, and the values of validated pointers.
Previous work has noted that LLVM's forward-edge CFI
implementation~\cite{CFI:Clang} may spill CFI sensitive data
to the stack~\cite{LoseControl:CCS15,ClangCFIBypass:GitHub}, making
the this data vulnerable to memory safety attacks.
{\ThisSystem} adds a new {\tt nospill} attribute to LLVM~IR, which will
prevent a virtual register from being spilled to the stack.
\begin{arXivOnly}
We discuss the details of {\tt nospill} in \Cref{appendix:nospill}.
\end{arXivOnly}
\begin{IEEEOnly}
We discuss the details of {\tt nospill} in our technical report~\cite{richter2024DeTRAParXiv}.
\end{IEEEOnly}
\subsection{Code Scanner}
\label{sec:impl-scanner}

We implemented the {\ThisSystem} code scanner, discussed
in \Cref{sec:design-scanner}, using LLVM's MC disassembler library.
The scanner first identifies all reachable code by examining the symbol table
for all functions, including forward-edge CFI jumptable entries
(see \Cref{sec:impl-compiler}) and \texttt{switch} jumptable destination
pointers.
It also reads the section headers to be able to distinguish between trusted and
untrusted code.

Next, inspired by the static analyzer of
Jalyoan, et. al.~\cite{ROPonRISCV:ASIACCS2020}, the scanner traces all possible
execution paths, generating a directed graph of basic blocks.
The input value for return instructions
(\mbox{\ASMInstruction{jr} \ASMRegister{ra}})
is checked to ensure that it is either unmodified since the preceding
\ASMInstruction{call}, or was loaded from the shadow stack.

To handle connecting basic blocks that are linked via indirect branches, the
scanner checks the instructions leading up to the jump to ensure that the
destination is either statically known (e.g., a long jump, which first requires an
\ASMInstruction{auipc} [add upper immediate to PC] instruction before the
\ASMInstruction{jr} [jump to offset from register] jump) or is loaded from
(\texttt{switch}) or checked against (indirect call) a jumptable, and the
results are used when connecting the basic blocks.
Whitelists were added for handwritten assembly that performs safe indirect jumps
that do not rely on jumptables (e.g., newlib's \mbox{RISC-V}
\texttt{memset}, which includes jumps whose offset is directly computed, rather
than being loaded from a table).

While tracing the discovered instructions in untrusted code, the scanner checks
for corruption of the shadow stack pointer, and other dangerous
instructions (see \Cref{sec:design-scanner}).

\subsection{Runtime}
\label{sec:impl-runtime}

Applications are linked to a custom runtime that contains trusted code,
including startup code necessary to implement {\ThisSystem}.
Standard library support is provided by the \mbox{RISC-V} newlib
port~\cite{riscv-newlib} based on revision \texttt{83d4bf},
with compiler support routines from compiler-rt \mbox{15.0.7}.
The runtime also includes code for tracking and reporting the outputs
of performance counters via the serial port.

\section{Security Benefit}
\label{sec:security}

To examine {\ThisSystem}'s security, we examined its effectiveness
against RiscyROP~\cite{RiscyROP:RAID2022}: the most sophisticated
attack against RISC-V of which we know.
RiscyROP found that, compared to x86 and ARM32,
it is more challenging to find useful gadgets on RISC-V,
due to multiple factors such as differences in calling conventions.
However, it also found that one can launch powerful code-reuse attacks to
call arbitrary functions with attacker-controlled arguments.
RiscyROP mainly exploits two types of gadgets:
(1) those that load a return address from the stack and return to that
address (called stack-based jump),
and (2) those that jump to an attacker-controlled register
(called jump-to-register), which corresponds to indirect function calls.
RiscyROP analyzed {\tt libc} and several applications and reported that
the majority of gadgets are stack-based
jumps,\footnote{Figure~3 of RiscyROP~\cite{RiscyROP:RAID2022} shows the
distribution of gadgets, but the paper does not summarize the statistics.}
which are also used multiple
times in its proof-of-concept attack.
{\ThisSystem} provides RAI, thus preventing stack-based jump gadgets from
being exploited.
As a result, {\ThisSystem} can thwart the attack demonstrated by RiscyROP.
Additionally, RiscyROP's jump-to-register gadgets can be
exploited to target arbitrary locations, while {\ThisSystem} restricts
those gadgets to only target the beginning of a group of functions
using CFI (\Cref{sec:design-compiler-forward-CFI}), which also mitigates
Jump-Oriented Programming~\cite{JOP:ASIACCS11}.
Overall, {\ThisSystem} significantly reduces the control-flow hijacking
attack surface for RISC-V.

\section{Performance Evaluation}
\label{sec:performance}

\begin{table}
	\centering
    \caption{Generated System-on-Chip Configuration}
    \begin{tabular*}{\columnwidth}{@{\extracolsep{\fill}}p{9.5em}|c}
        \hline
        Core                  & rv32imafdc at 50 MHz              \\
        Branch Target Buffer  & 28-entry                          \\
        Branch History Table  & 512-entry                         \\
        Return Address Stack  & 6-entry                           \\
        Breakpoints           & 8, address match only             \\
        Phys Mem Protection   & 8 regions, 4 byte granularity     \\
        Cache line            & 64 bytes                          \\
        L1 Data               & 64~KiB, 4-way                     \\
        L1 Code               & 16~KiB, 2-way                     \\
        L2 (Shared Inclusive) & 256~KiB, 8-way{\quad}5 MSHRs      \\
        On-board DDR3         & 256 MiB $\times$16 at 333 MHz CL5 \\
        \hline
    \end{tabular*}
    \label{table:fpgaconfig}
\end{table}

To evaluate {\ThisSystem}'s performance, we used the Chipyard~\cite{Chipyard}
System-on-Chip (SoC) framework version 1.6.2 to generate verilog for a full
system with our modified
Rocket core~\cite{RocketChip} \mbox{RISC-V} implementation.
We ran our design on a Digilent Arty A7-100T Development
Board~\cite{Arty-A7:Digilent} and used Xilinx Vivado 2021.2 to synthesize and
implement the verilog to run on the on-board 
\mbox{XC7A100TCSG324-1} FPGA.

We configured the SoC to be similar to the SiFive Freedom
E310~\cite{E310-Arty:SiFive} Arty (an E31~\cite{E31:SiFive} core implementation
for Arty A7 development boards) and FE310~\cite{FE310:SiFive} SoC.
To support large applications that require more memory than available on the
FPGA, such as those in CoreMark Pro~\cite{CoreMarkPro:EEMBC},
we changed the memory system to be backed by the 256~MiB on-board DRAM, and
use a 256~KiB shared inclusive L2 cache sized to fit in the remaining FPGA SRAM.
The L1 code and data caches are 16~KiB and 64~KiB, respectively, corresponding
to the sizes of the code and data tightly-integrated memories on the (F)E310.
The L1 data cache is 4-way set associative (the same as the L1 data caches on an
ARM M7~\cite{ARMCortexM7PTR} or M55~\cite{ARMCortexM55PTR}). %
Like in the E310, our L1 code cache is 2-way set associative but lacks
the tightly-integrated memory (ITIM) functionality.
To evaluate benchmarks that use hardware floating-point instructions, 
we added a 64-bit FPU to the core
(the FPU is an optional feature on E31 cores~\cite{E31:SiFive}).
We also increased the number of additional event counters from two to the ISA
maximum of 29 for enhanced data collection.
Table~\ref{table:fpgaconfig} shows the full configuration.

\subsection{Build Configuration}
\label{sec:performance-builds}

We used our modified Clang/\mbox{LLVM} toolchain and runtime
(see \Cref{sec:impl-compiler,sec:impl-runtime}) to build  the evaluated
benchmarks.
We compiled all code with \verb|-O2| optimizations, link-time optimization, and
linker relaxation enabled.
We compiled benchmarks to use
hardware floating-point instructions and the floating-point ABI.
Except for any
file-specific or benchmark-specific flags, all compiler and linker options,
including those for optimization and sanitizers, are common across all sources
(including compiler-rt).
We compared this baseline to a binary that additionally enables {\ThisSystem}
protections. %

\subsection{Code Layout Effects on Performance}
\label{sec:performance-layout}

When evaluating benchmark performance, we observed run-to-run
performance variations of less than 0.1\%.
However, as also seen in previous work~\cite{mytkowicz2009producing},
changes in memory layout impacted
performance by 1\% or more.
For example, a build with all {\ThisSystem} protections enabled could execute
faster than one with none of its protections, even though the {\ThisSystem}
build executes more instructions.

To reduce the chance that fortuitous memory layouts make our
approach faster,
we compiled each benchmark with 100 different layouts: one
layout generated by the default settings in the compiler and linker and 99
pseudorandom layouts provided by LLD's \verb|--shuffle-sections|
option~\cite{LLDShuffleSections}.
We then report results from the fastest layout of each build.
The fastest layouts of each build are compared to each other even though
they are
likely to have been linked with different \verb|--shuffle-sections| values.
When evaluating generated code sizes, we also use the sizes for the fastest
build.

\subsection{Benchmark Suites}
\label{sec:performance-applications}

We evaluated several benchmarks.
\textbf{CoreMark-Pro}~\cite{CoreMarkPro:EEMBC} benchmarks are from revision
\texttt{4832cc}.  We set the iteration count for each benchmark to the smallest
value that would still run for at least 10~seconds.
We modified the zip benchmark to use pre-computed sample data, rather than
generating it on-chip during the untimed initialization phase.
\textbf{Embench}~\cite{embench:riscv2019} benchmarks are from revision \texttt{d9b30c}.  We left the iteration counts unmodified.
The number of iterations of each benchmark was based on their
scaling factors divided by our processor's speed.
\textbf{BEEBS}~\cite{BEEBS:arXiv} benchmarks were obtained from its git
repository~\cite{BEEBS:github}, commit \texttt{049ded}.
We set each benchmark's iteration count so that it would run for at least 1
second or 100~iterations, whichever was longer.
\textbf{CoreMark}~\cite{CoreMark:EEMBC} is from revision \texttt{b24e397}
and used unmodified.

For some programs in BEEBS and Embench, the compiler was able to optimize away the entire
benchmark's computation.  In several cases, constant propagation allowed the
compiler to calculate the benchmark's result at compile-time, so the compiler
transformed the benchmark to simply output
the precomputed result.  For other benchmarks, the result was never used,
so the compiler removed the computation altogether as unnecessary.
Alternately, the compiler determined that each iteration performed an
identical computation, and so emitted code that only performed the computation
once regardless of how many iterations were requested.
Affected Embench benchmarks were identified by manually examining the generated
native code for benchmarks that ran for less than one second.
For BEEBS benchmarks, we ran each benchmark with varying iteration
counts, using the instruction retire count from each run to establish how many
instructions were run per iteration.  We then manually examined the
generated native code for benchmarks that ran fewer than 100 instructions per
iteration.
Once identified, we added empty inline assembly statements to prevent these
optimizations from removing the benchmark's core computation.
Inputs were marked as ``written'' at the beginning of each iteration
and outputs as ``read''. %
Embench benchmarks modified to prevent these optimizations were: cubic,
st, statemate, and tarfind. BEEBS benchmarks modified were: aha-compress, bs,
crc, crc32, cubic, fibcall, frac, janne, lcdnum, nbody,
newlib-exp, newlib-mod, ns, qurt, sglib-queue and whetstone.
Floating-point benchmarks that performed verification against expected values --
ludcmp, matmult, nbody, st, and stb\_perlin --
were modified to allow for a small difference in the expected result due to
floating-point rounding differences.
We also fixed out-of-bounds array accesses in select and qsort,
corrected duff to use the correct source and destination arrays,
and modified function parameter types in sha256 to work properly with
LLVM's indirect function call type checking.

\subsection{Execution Times}
\label{sec:performance-performance}

\setlength{\tabcolsep}{4pt}
\begin{table}[tb]
\caption{Execution Times}
\sffamily
\centering
\begin{tabular}[t]{lrrlrr}
\hhline{======}
\textbf{Benchmark} & \textbf{-O2} & \textbf{{\ThisSystem}} & \textbf{Benchmark} & \textbf{-O2} & \textbf{{\ThisSystem}}\\
 & \textbf{(s)} & ($\mathbf{\times}$) &  & \textbf{(s)} & ($\mathbf{\times}$)\\
\hhline{======}
\multicolumn{6}{c}{\textbf{CoreMark-Pro}} \\
\hline
\textbf{cjpeg} & 10.51 & 1.003 & \textbf{parser} & 11.74 & 1.006 \\
\textbf{core} & 191.1 & 1.031 & \textbf{radix} & 10.47 & 1.000 \\
\textbf{linear} & 12.20 & 1.000 & \textbf{sha} & 10.18 & 1.006 \\
\textbf{loops} & 51.34 & 1.002 & \textbf{zip} & 10.53 & 1.000 \\
\textbf{nnet} & 49.36 & 1.001 &  &  & \\
\hline
\textbf{min} & 10.18 & 1.000 &  &  &  \\
\textbf{max} & 191.1 & 1.031 &  &  &  \\
\textbf{geomean} &  & 1.005 &  &  &  \\
\hhline{======}
\multicolumn{6}{c}{\textbf{Embench}} \\
\hline
\textbf{aes} & 3.723 & 1.000 & \textbf{picojpeg} & 4.326 & 1.012 \\
\textbf{crc32} & 2.961 & 1.000 & \textbf{primecount} & 12.45 & 1.000 \\
\textbf{cubic} & 2.077 & 1.017 & \textbf{qrduino} & 3.332 & 1.001 \\
\textbf{edn} & 6.902 & 1.000 & \textbf{sglib} & 3.285 & 1.011 \\
\textbf{huffbench} & 2.872 & 0.999 & \textbf{sha256} & 3.834 & 1.035 \\
\textbf{matmult-int} & 2.758 & 1.055 & \textbf{slre} & 3.464 & 1.013 \\
\textbf{md5sum} & 2.347 & 0.998 & \textbf{st} & 0.184 & 1.000 \\
\textbf{minver} & 0.518 & 1.000 & \textbf{statemate} & 0.216 & 1.018 \\
\textbf{mont64} & 5.716 & 1.006 & \textbf{tarfind} & 1.586 & 1.001 \\
\textbf{nbody} & 0.180 & 1.000 & \textbf{ud} & 3.834 & 1.002 \\
\textbf{nsichneu} & 3.097 & 1.016 & \textbf{wikisort} & 0.399 & 1.133 \\
\hline
\textbf{min} & 0.180 & 0.998 &  &  &  \\
\textbf{max} & 12.45 & 1.133 &  &  &  \\
\textbf{geomean} &  & 1.014 &  &  &  \\
\hhline{======}
\multicolumn{3}{c}{\textbf{BEEBS} (Summary)} & \multicolumn{3}{c}{\textbf{CoreMark}}\\ %
\hline
\textbf{min} & 1.015 & 0.991 &  &  &  \\
\textbf{max} & 1.489 & 1.201 & \textbf{coremark} & 10.40 & 1.019 \\ %
\textbf{geomean} &  & 1.010 &  &  &  \\
\hhline{======}
\end{tabular}
\label{table:all-performance}
\end{table}
\setlength{\tabcolsep}{6pt}

Table~\ref{table:all-performance} shows runtime performance for the
CoreMark-Pro, Embench, and CoreMark benchmarks.
Due to limited space, Table~\ref{table:all-performance} only shows a
statistical summary of the 80 individual benchmarks in BEEBS; full results
\begin{arXivOnly}
from BEEBS can be found in \Cref{appendix:beebs}.
\end{arXivOnly}
\begin{IEEEOnly}
from BEEBS can be found in our technical report~\cite{richter2024DeTRAParXiv}.
\end{IEEEOnly}
Across all benchmarks we evaluated, the relative {\ThisSystem} performance
ranged from \mbox{0.991$\times$} (0.9\% \textit{faster}) to
\mbox{1.201$\times$} (20.1\% slower).
The geometric mean across the 112 individual benchmarks was
\mbox{1.011$\times$} (1.1\% overhead).

\textbf{Comparing to Related Work.\quad}
For the subset of BEEBS benchmarks reported by Silhouette~\cite{Silhouette:UsenixSec20},
{\ThisSystem} performance ranged from \mbox{0.991$\times$} to \mbox{1.131$\times$}, with a
geometric mean of \mbox{1.008$\times$}.
By comparison, Silhouette's performance overhead on these benchmarks was between
\mbox{1.001$\times$} and \mbox{1.510$\times$}, averaging \mbox{1.102$\times$}
--- {\ThisSystem} is 8.5\% faster. %
Silhouette also evaluated CoreMark Pro, with performance between \mbox{1.001$\times$}
and \mbox{1.049$\times$}, averaging \mbox{1.010$\times$}
--- {\ThisSystem}
 is 0.5\% faster. %
Although we did not evaluate against most benchmarks evaluated by
{\uRAI}~\cite{uRAI:NDSS20}, we did run the CoreMark
benchmark~\cite{CoreMark:EEMBC}.
{\ThisSystem}'s overhead on CoreMark is 1.9\%, while {\uRAI}'s overhead is 8.1\%
--- {\ThisSystem} is 5.7\% faster than {\uRAI}.

\subsection{Code Size}
\label{sec:performance-sizing}

Embedded systems often have limited memory; keeping code size small is critical.
We therefore evaluated {\ThisSystem}'s code size overheads by measuring the
size of the code sections of each ELF executable.
For each build of each benchmark, we measured size from the memory
layout that had the smallest execution time.
Due to limited space, we summarize the results in
\Cref{table:all-sizes-summary};
\begin{arXivOnly}
full results can be seen in \Cref{appendix:sizes}.
\end{arXivOnly}
\begin{IEEEOnly}
full results can be seen in our technical report~\cite{richter2024DeTRAParXiv}.
\end{IEEEOnly}

{\ThisSystem} has code size overheads that average 4.5\% across all benchmarks.
Compared to Silhouette, which had a code size overhead of 8.9\% on CoreMark
Pro and 2.3\% on a subset of BEEBS, {\ThisSystem}'s respective overheads are
7.9\% and 6.7\% --- 1\% better and 4.4\% worse.
However, we note that {\ThisSystem} instruments the standard library and
board support code, which Silhouette does not.
When we subtract the standard library code, Silhouette's average code size
overheads become 16.5\% on CoreMark Pro and 5.3\% on BEEBS, making
{\ThisSystem}'s overheads 7\% better and 1.3\% worse respectively.

Instead of evaluating code size, {\uRAI} reports ``Flash'' utilization,
which we understand to include code, read-only data, and initialized writable
data.
On CoreMark, {\uRAI} has $\mathrm{\sim}$40\% Flash overhead, while {\ThisSystem}'s
code, data, and rodata shows a 2.7\% reduction in size --- {\ThisSystem} is
$\mathrm{\sim}$40\% better.

\setlength{\tabcolsep}{5pt}
\begin{table}[bt]
\caption{Relative Code Sizes}
\sffamily
\centering
\begin{tabular}[t]{lrrlrr}
\hline
 & \textbf{-O2} & \textbf{{\ThisSystem}} & & \textbf{-O2} & \textbf{{\ThisSystem}}\\
 & \textbf{(KiB)} & ($\mathbf{\times}$) &  & \textbf{(KiB)} & ($\mathbf{\times}$)\\
\hline
\multicolumn{3}{c}{\textbf{CoreMark-Pro}} & \multicolumn{3}{c}{\textbf{Embench}} \\
\textbf{min} & 28.50 & 1.043 & \textbf{min} & 16.74 & 0.947 \\
\textbf{max} & 52.25 & 1.238 & \textbf{max} & 34.43 & 1.101 \\
\textbf{geomean} &  & 1.079 & \textbf{geomean} &  & 0.965 \\
\hline
\multicolumn{3}{c}{\textbf{BEEBS}} & \multicolumn{3}{c}{\textbf{CoreMark}} \\
\textbf{min} & 7.254 & 0.951 &  &  &  \\
\textbf{max} & 28.14 & 1.375 & \textbf{coremark} & 22.68 & 0.964 \\
\textbf{geomean} &  & 1.064 &  &  &  \\
\hline
\end{tabular}
\label{table:all-sizes-summary}
\end{table}\setlength{\tabcolsep}{6pt}

\subsection{Hardware Utilization}
\label{sec:performance-utilization}

We evaluated the increased chip area (a proxy for manufacturing cost)
needed to implement the required features for {\ThisSystem}
(see \Cref{sec:impl-rocket}).
We used Chipyard~\cite{Chipyard} version 1.10.0's Hammer~\cite{Hammer}
VLSI design flow, utilizing OpenROAD~\cite{OpenROAD} to implement
the design for the Sky130 PDK~\cite{edwards2020google}.

The baseline design was a TinyRocketConfig with 4 debug trigger registers.
We compared this to a modified 4-trigger design that meets {\ThisSystem}'s
requirements.
Our changes increase the unrouted pipeline core area
(not counting cache or scratchpad) by 0.14\%.
For comparison, the 20~KiB of SRAM arrays used by the cache and scratchpad
require 800\% more area than a routed pipeline core.

\section{Related Work}
\label{sec:related}

\JZ{The main point of writing related work is to demonstrate how the current
work differs, ideally betters, prior work. Every related work should be compared
to this work. There is no need to do comparison one by one---it's okay and
common to put several prior works in a group and compare this group of work
with this paper, but they should be compared. The current work does not
necessarily have to surpass prior work in every way. Admitting limitations
compared to prior work is fine.}

\JZ{``memory management hardware'' seems not to be a common term. It could be
a bit confusing without further explanation or examples. I'd suggest rephrase it.}

Memory management hardware has been used to
mitigate memory safety attacks, even within
single-address-space embedded applications.
Kage~\cite{Kage:UsenixSec22} and Silhouette~\cite{Silhouette:UsenixSec20}
utilize the ARM Memory Protection Unit (MPU) to create memory regions
that only normal store instructions can modify; they then
transform all untrusted store instructions into store-with-translation
instructions that cannot write into these protected regions.  They
place shadow stacks and other security-critical data in these protected regions.
uXOM~\cite{uXOM:USS19} uses the same technique to implement execute-only memory.

RECFISH~\cite{RECFISH:ecrts19} also uses a MPU-protected shadow stack but
requires a supervisor call to privileged code to push return addresses.
IskiOS~\cite{IskiOS:RAID21} leverages
Intel PKU~\cite{IntelArchManual21} to secure a shadow stack, temporarily
enabling writes to the stack via a configuration register while
writing a return address.
CHERI~\cite{CHERI:ISCA14} uses hardware-enforced capabilities, ensuring that
new capabilities can be derived only from preexisting capabilities.
Even if an application overwrites a return address, unless that
write happened to be a valid code pointer, attempts to return to the corrupted
address will fail.
RetTag~\cite{RetTag:EuroSec2022} adds pointer authentication instructions to the
\mbox{RISC-V} ISA to authenticate return addresses.  In contrast,
{\ThisSystem} makes no ISA modifications,
requiring only the existing debug ISA.

\uRAI~\cite{uRAI:NDSS20} statically computes a complete call
graph and encodes all jumps statically in read-only code jumptables.
A dedicated register encodes the current location on the call tree, allowing
the code to check against each possible return location, and return
specifically to that location.
If the register is corrupted, the jumptable lookup will not find a valid
return address and fail.
{\ThisSystem} does not need to compute a
complete callgraph and has less code memory overhead.

\mbox{O-CFI}~\cite{mohan2015OCFI} uses layout modification to ensure that all
valid
indirect targets have a known alignment, and clustering, which allows
a bounds check to determine the validity of a branch target.
The location of the table of valid bounds is randomized and saved only to a
register, preventing leaks of the bounds lookup table's (BLT's) base address.
However, \mbox{O-CFI}'s protections will fail if an attacker can find the
table, for example via a side channel attack of the BLT register saved to a
kernel stack, or by scanning read-only memory locations for values that match
a known valid indirect pointer (e.g. a return address spilled to the stack).
Redactor~\cite{Redactor:Oakland15} uses execute-only memory and statically
generated trampolines with random memory and register layouts to prevent
an attacker from reliably generating a usable gadget chain.
Moreover, unlike {\ThisSystem}, both \mbox{O-CFI} and Redactor must compute a
reverse control-flow graph ahead of time, and their reverse CFI is not
context-sensitive; an
attacker can potentially redirect reverse control flow to the
wrong caller, even without knowledge of the BLT's location.
{\ThisSystem} provides context-sensitive reverse CFI, even against
omniscient attackers.

Work most closely related to ours has used debugging hardware to enforce
non-discriminatory security policies that apply to all code; once
configured, enabling access requires explicitly disabling the watchpoint.
PicoXOM~\cite{PicoXOM:SecDev20} provides execute-only code memory
using watchpoint hardware and
prevents reconfiguration of the memory-mapped watchpoint registers.
Jang, et. al.~\cite{JangInProcessSecurity:DAC19} used
watchpoints to allow an application to selectively lock and unlock
regions via a system call;
they also use watchpoints to prevent kernel access to
user memory and to make kernel memory execute-only~\cite{JangPAN:DAC19}.

PHMon~\cite{PHMon:UsenixSec20} adds an execution trace/monitor unit to the
processor core.
This monitor includes a small programmable unit that can perform actions in
response to detected events.
One use of the unit is to implement its own shadow stack, listening for call
and return instructions to know when to push and pop addresses, interrupting
the system if it detects a return to an address that does not match what it
saved.
Compared with {\ThisSystem}, PHMon adds substantial hardware to the core, and
because it only monitors a trace of completed execution, it can only throw a
trap after instructions have been committed.

\section{Future Work}
\label{sec:future}

Several directions exist for future work.
We can use debug triggers to solve other security
challenges, such as protecting additional control data~\cite{Kage:UsenixSec22}
or isolating application components~\cite{Hodor:Usenix19}.
We can also explore whether improvements to debug triggers
e.g., new data matching features or longer chain support,
improves their utility for security enforcement.

\section{Conclusion}
\label{sec:conclusion}

We presented {\ThisSystem} which provides embedded
applications with return address integrity by utilizing
\mbox{RISC-V} debug facilities,
novel compiler transformations,
and a trusted runtime
to protect a shadow stack from corruption.

\begin{arXivOnly}
\section*{Availability}
\end{arXivOnly}
{\ThisSystem}'s source code is available from
\hyperlink{https://github.com/URSec/DeTRAP}{https://github.com/URSec/DeTRAP}.

\begin{arXivOnly}
\section*{Acknowledgments}
\end{arXivOnly}
This work was supported by NSF Grants CNS 1652280 and CNS 2154322.
The authors gratefully acknowledge Komail Dharsee for the idea of using debug
triggers to implement a security policy.

\bibliographystyle{IEEEtranS}
\begin{IEEEOnly}
\bibliography{\jobname}
\end{IEEEOnly}
\begin{arXivOnly}
\bibliography{\jobname,FixDashes\jobname.bib}
\end{arXivOnly}

\begin{arXivOnly}

\appendices
\label{sec:appendix}

\section{\texttt{setjmp} and \texttt{longjmp} handling}
\label[appendix]{appendix:setjmp}

Normally, the forward-edge CFI prevents shadow stack underflow by ensuring that
calls
cannot be made to the function's epilogue unless its trampoline was previously
called, and that every execution of the trampoline is eventually matched by
execution of the epilogue.
However, 
improper use of \texttt{setjmp} and \texttt{longjmp} functions could cause
imbalanced execution of the trampolines and epilogues.
\texttt{longjmp} restores processor state from memory, including the
shadow stack pointer, potentially restoring
corrupted
values in violation of \Cref{inv:ssread}.
Although use of such functions on embedded systems is rare --- none of our
evaluated benchmarks utilize them --- 
{\ThisSystem} nonetheless adopts the methodology of Silhouette~\cite{Silhouette:UsenixSec20}
to safely store processor state in a \textbf{trusted map} within the
write-limited region, indexed by the \texttt{jmp\_buf} argument to \texttt{setjmp}.
When calling \texttt{longjmp}, \texttt{jmp\_buf} must refer to a valid entry in
the trusted map; otherwise, an exception is raised, and {\ThisSystem} acts as if
untrusted code attempted to modify the write-limited region.

Just before \texttt{longjmp} resumes execution at the point saved by
\texttt{setjmp}, it examines the trusted map and
invalidates any entry that refers to now-stale stack frames, identified by
a shadow stack pointer value that is greater than the current shadow stack
pointer value.
This ensures that a future \texttt{longjmp} call cannot restore the saved state
of a callee function that is now unwound.
For the same reason, when a function contains a call to \texttt{setjmp}, the
compiler instruments its
epilogue to add a call to trusted code that invalidates any trusted
map entries created while the function was executing.

\section{{\tt nospill} Attribute for CFI-Sensitive Data}
\label[appendix]{appendix:nospill}

Clang's forward CFI~\cite{CFI:Clang} relies on the output of the
LLVM~IR \texttt{llvm.type.test} intrinsic.
Early in compilation, the compiler lowers this intrinsic to IR primitives that
load relevant constants and determine if the pointer is valid.
However, because the lowering happens early in compilation,
the generated code is subject to various optimizations,
especially loop-invariant code motion and constant hoisting.
These optimizations can cause CFI sensitive data to be spilled onto the stack,
leaving them vulnerable
to memory corruption~\cite{LoseControl:CCS15,liebchen2018thesis}.
Similar vulnerabilities also exist when a switch statement is implemented with
a jumptable: constants with the jumptable address and bounds as well as the
code address loaded from the jumptable can all be spilled to the stack.

To prevent such vulnerabilities in our prototype, we added a new
\texttt{nospill} attribute to the LLVM~IR.  When placed on an SSA
virtual register, the LLVM compiler will prevent the value from being
spilled to the untrusted stack.  Instead, if the value needs to be
reloaded into a register, it will be reloaded from read-only memory or
rematerialized via load immediate/load address instructions.  We then
modified the LLVM CFI implementation to place the {\tt nospill}
attribute on all CFI sensitive data.  With this attribute, our
prototype permits the compiler to safely optimize CFI sensitive
data, including performing constant hoisting, so long as the compiler can
ensure that the register holding the data will not be spilled to the stack.
For example, if none of the potential indirect callee functions need to save
any callee-save registers, the calling function can keep CFI sensitive data in
those registers across the call.

While simple in theory, adding the {\tt nospill} attribute posed a
significant engineering challenge.  In most cases, the libraries that
the compiler uses to modify LLVM~IR do not allow a value to be
changed; rather, the compiler creates a new value and replaces all
uses of the old value with the new value.  Consequently, optimizations
may inadvertently strip the {\tt nospill} attribute from a value.
We therefore built a binary analysis tool which checks whether CFI
sensitive data is spilled to the stack.  Using this tool, we verified
that our {\tt nospill} attribute prevents unsafe spills for all
benchmarks except for CoreMark~Pro's {\tt cjpeg}; for {\tt cjpeg},
there are three indirect calls in which the register holding the size of
the jumptable is loaded from the stack.  This appears to be a bug in
the compiler as, on RISC-V, rematerializing these integer constants via a
load-immediate instruction (because they are all less than $15$) should be
preferred to loading it from the stack.  We expect that future
versions of LLVM will improve code generation and fix this bug.

A complete implementation of {\tt nospill} would include the analyses necessary
to determine if an optimization of a {\tt nospill} value was safe.
As {\tt nospill} is not a focus of this paper, we did not implement those
analyses, and instead just assumed that all optimizations were unsafe ---
our implementation assumes that all {\tt nospill} values need to be
recreated and/or checked at each use.

\section{BEEBS Execution Times}
\label[appendix]{appendix:beebs}

\Cref{table:beebs-performance} shows the execution times for each benchmark
in BEEBS. The statistical summary (min/max/geomean) is in
\Cref{table:all-performance}.
\setlength{\tabcolsep}{4pt}
\begin{table}[h]
\footnotesize
\sffamily
\centering
\caption{BEEBS Execution Times}
\begin{tabular}[t]{lrrlrr}
\hline
\textbf{Benchmark} & \textbf{-O2} & \textbf{{\ThisSystem}} & \textbf{Benchmark} & \textbf{-O2} & \textbf{{\ThisSystem}}\\
    & \textbf{(s)} & ($\mathbf{\times}$) &  & \textbf{(s)} & ($\mathbf{\times}$)\\
\hline
\textbf{aes} & 1.093 & 1.003 & \textbf{nettle-cast128} & 1.094 & 1.007 \\
\textbf{aha-compress} & 1.092 & 1.000 & \textbf{nettle-des} & 1.100 & 0.999 \\
\textbf{bs} & 1.047 & 1.000 & \textbf{nettle-md5} & 1.100 & 1.000 \\
\textbf{bubblesort} & 1.100 & 1.000 & \textbf{nl-exp} & 1.041 & 1.006 \\
\textbf{cnt} & 1.094 & 1.000 & \textbf{nl-log} & 1.086 & 1.005 \\
\textbf{compress} & 1.049 & 1.040 & \textbf{nl-mod} & 1.067 & 1.031 \\
\textbf{cover} & 1.097 & 1.000 & \textbf{nl-sqrt} & 1.087 & 0.999 \\
\textbf{crc} & 1.099 & 1.003 & \textbf{ns} & 1.088 & 1.000 \\
\textbf{crc32} & 1.100 & 1.000 & \textbf{nsichneu} & 1.076 & 1.004 \\
\textbf{ctl-stack} & 1.099 & 1.000 & \textbf{picojpeg} & 1.401 & 1.012 \\
\textbf{ctl-string} & 1.034 & 1.005 & \textbf{prime} & 1.100 & 1.000 \\
\textbf{ctl-vector} & 1.088 & 1.000 & \textbf{qrduino} & 1.335 & 0.997 \\
\textbf{cubic} & 1.027 & 1.018 & \textbf{qsort} & 1.078 & 0.999 \\
\textbf{dijkstra} & 1.223 & 1.000 & \textbf{qurt} & 1.101 & 1.004 \\
\textbf{dtoa} & 1.057 & 1.029 & \textbf{recursion} & 1.098 & 1.201 \\
\textbf{duff} & 1.102 & 1.000 & \textbf{rijndael} & 1.489 & 1.000 \\
\textbf{edn} & 1.083 & 1.001 & \textbf{select} & 1.082 & 1.000 \\
\textbf{expint} & 1.100 & 1.000 & \textbf{sg-array} & 1.045 & 1.000 \\
\textbf{fac} & 1.089 & 1.000 & \textbf{sg-arrheap} & 1.095 & 1.001 \\
\textbf{fasta} & 1.100 & 1.000 & \textbf{sg-arrquick} & 1.096 & 1.000 \\
\textbf{fdct} & 1.088 & 1.003 & \textbf{sg-dllist} & 1.098 & 1.000 \\
\textbf{fibcall} & 1.100 & 1.000 & \textbf{sg-hashtable} & 1.097 & 1.001 \\
\textbf{fir} & 1.101 & 1.000 & \textbf{sg-listinsert} & 1.098 & 1.000 \\
\textbf{frac} & 1.100 & 0.998 & \textbf{sg-listsort} & 1.100 & 1.000 \\
\textbf{huffbench} & 1.096 & 0.999 & \textbf{sg-queue} & 1.087 & 1.000 \\
\textbf{insertsort} & 1.087 & 1.000 & \textbf{sg-rbtree} & 1.069 & 0.991 \\
\textbf{janne\_complex} & 1.100 & 1.000 & \textbf{sha256} & 1.087 & 1.013 \\
\textbf{jfdctint} & 1.093 & 1.002 & \textbf{slre} & 1.056 & 1.020 \\
\textbf{lcdnum} & 1.092 & 1.000 & \textbf{sqrt} & 1.101 & 1.000 \\
\textbf{levenshtein} & 1.105 & 1.008 & \textbf{st} & 1.071 & 1.000 \\
\textbf{ludcmp} & 1.100 & 0.995 & \textbf{statemate} & 1.066 & 1.030 \\
\textbf{matmult-float} & 1.100 & 1.000 & \textbf{stb\_perlin} & 1.100 & 1.000 \\
\textbf{matmult-int} & 1.055 & 1.046 & \textbf{stringsearch1} & 1.074 & 0.996 \\
\textbf{mergesort} & 1.287 & 1.095 & \textbf{strstr} & 1.015 & 1.000 \\
\textbf{miniz} & 1.079 & 1.017 & \textbf{tarai} & 1.100 & 1.149 \\
\textbf{minver} & 1.072 & 1.000 & \textbf{trio-snprintf} & 1.096 & 0.994 \\
\textbf{mont64} & 1.096 & 1.001 & \textbf{trio-sscanf} & 1.075 & 1.003 \\
\textbf{nbody} & 1.102 & 1.000 & \textbf{ud} & 1.100 & 1.000 \\
\textbf{ndes} & 1.091 & 1.000 & \textbf{whetstone} & 1.095 & 1.011 \\
\textbf{nettle-arcfour} & 1.100 & 1.000 & \textbf{wikisort} & 1.218 & 1.131 \\
\hline
\end{tabular}
\label{table:beebs-performance}
\end{table}\setlength{\tabcolsep}{6pt}

\section{Code Sizes}
\label[appendix]{appendix:sizes}

\Cref{table:all-sizes} shows the code sizes for the individual benchmarks
in each suite.
Note that CoreMark only has one benchmark, and its result
is in \Cref{table:all-sizes-summary}, along with the statistical summary
(min/max/geomean) for each suite.

\setlength{\tabcolsep}{4pt}
\begin{table}[h]
\caption{Relative Code Sizes}
\sffamily
\centering
\begin{tabular}[t]{lrrlrr}
\hline
\textbf{Benchmark} & \textbf{-O2} & \textbf{{\ThisSystem}} & \textbf{Benchmark} & \textbf{-O2} & \textbf{{\ThisSystem}}\\
 & \textbf{(KiB)} & ($\mathbf{\times}$) &  & \textbf{(KiB)} & ($\mathbf{\times}$)\\
\hline
\multicolumn{6}{c}{\textbf{CoreMark-Pro}} \\
\textbf{cjpeg} & 52.25 & 1.238 & \textbf{parser} & 36.03 & 1.063 \\
\textbf{core} & 28.50 & 1.074 & \textbf{radix} & 32.97 & 1.062 \\
\textbf{linear} & 31.10 & 1.065 & \textbf{sha} & 37.26 & 1.051 \\
\textbf{loops} & 44.41 & 1.063 & \textbf{zip} & 51.94 & 1.043 \\
\textbf{nnet} & 32.84 & 1.061 &  &  & \\
\hline
\multicolumn{6}{c}{\textbf{Embench}} \\
\textbf{aes} & 19.06 & 0.956 & \textbf{picojpeg} & 34.43 & 1.018 \\
\textbf{crc32} & 16.74 & 0.950 & \textbf{primecount} & 16.89 & 0.947 \\
\textbf{cubic} & 32.85 & 0.978 & \textbf{qrduino} & 26.48 & 0.968 \\
\textbf{edn} & 17.92 & 0.955 & \textbf{sglib} & 19.25 & 0.953 \\
\textbf{huffbench} & 18.28 & 0.953 & \textbf{sha256} & 21.46 & 0.964 \\
\textbf{matmult-int} & 17.40 & 0.952 & \textbf{slre} & 20.08 & 0.959 \\
\textbf{md5sum} & 17.43 & 0.952 & \textbf{st} & 17.23 & 0.951 \\
\textbf{minver} & 18.15 & 0.953 & \textbf{statemate} & 17.27 & 0.952 \\
\textbf{mont64} & 19.05 & 0.954 & \textbf{tarfind} & 17.14 & 0.951 \\
\textbf{nbody} & 17.38 & 0.948 & \textbf{ud} & 18.11 & 0.954 \\
\textbf{nsichneu} & 27.22 & 0.968 & \textbf{wikisort} & 18.25 & 1.101 \\
\hline
\multicolumn{6}{c}{\textbf{BEEBS}} \\
\textbf{aes} & 9.498 & 1.060 & \textbf{nettle-cast128} & 10.95 & 1.049 \\
\textbf{aha-compress} & 8.035 & 1.062 & \textbf{nettle-des} & 11.57 & 1.040 \\
\textbf{bs} & 7.299 & 1.069 & \textbf{nettle-md5} & 9.523 & 1.056 \\
\textbf{bubblesort} & 7.426 & 1.067 & \textbf{nl-exp} & 7.834 & 1.068 \\
\textbf{cnt} & 8.117 & 1.064 & \textbf{nl-log} & 7.814 & 1.069 \\
\textbf{compress} & 7.930 & 1.070 & \textbf{nl-mod} & 7.686 & 1.071 \\
\textbf{cover} & 7.291 & 1.073 & \textbf{nl-sqrt} & 8.887 & 1.058 \\
\textbf{crc} & 7.434 & 1.073 & \textbf{ns} & 7.557 & 1.068 \\
\textbf{crc32} & 7.307 & 1.071 & \textbf{nsichneu} & 17.29 & 1.029 \\
\textbf{ctl-stack} & 7.889 & 1.066 & \textbf{picojpeg} & 24.46 & 1.082 \\
\textbf{ctl-string} & 8.689 & 1.063 & \textbf{prime} & 7.408 & 1.069 \\
\textbf{ctl-vector} & 8.766 & 1.060 & \textbf{qrduino} & 16.91 & 1.033 \\
\textbf{cubic} & 28.14 & 1.026 & \textbf{qsort} & 7.857 & 1.064 \\
\textbf{dijkstra} & 8.066 & 1.067 & \textbf{qurt} & 7.629 & 1.070 \\
\textbf{dtoa} & 15.43 & 1.045 & \textbf{recursion} & 7.328 & 1.074 \\
\textbf{duff} & 7.672 & 1.067 & \textbf{rijndael} & 19.71 & 1.026 \\
\textbf{edn} & 8.391 & 1.061 & \textbf{select} & 7.586 & 1.069 \\
\textbf{expint} & 7.439 & 1.067 & \textbf{sg-array} & 7.344 & 1.070 \\
\textbf{fac} & 7.254 & 1.075 & \textbf{sg-arrheap} & 7.500 & 1.073 \\
\textbf{fasta} & 7.721 & 1.068 & \textbf{sg-arrquick} & 7.588 & 1.074 \\
\textbf{fdct} & 7.963 & 1.068 & \textbf{sg-dllist} & 7.672 & 1.069 \\
\textbf{fibcall} & 7.285 & 1.070 & \textbf{sg-hashtable} & 7.932 & 1.068 \\
\textbf{fir} & 7.389 & 1.071 & \textbf{sg-listinsert} & 7.373 & 1.074 \\
\textbf{frac} & 7.529 & 1.069 & \textbf{sg-listsort} & 7.578 & 1.068 \\
\textbf{huffbench} & 8.912 & 1.062 & \textbf{sg-queue} & 7.549 & 1.069 \\
\textbf{insertsort} & 7.506 & 1.069 & \textbf{sg-rbtree} & 8.113 & 1.065 \\
\textbf{janne\_complex} & 7.371 & 1.069 & \textbf{sha256} & 11.93 & 1.050 \\
\textbf{jfdctint} & 7.982 & 1.066 & \textbf{slre} & 10.57 & 1.056 \\
\textbf{lcdnum} & 7.346 & 1.070 & \textbf{sqrt} & 7.389 & 1.070 \\
\textbf{levenshtein} & 7.715 & 1.073 & \textbf{st} & 7.742 & 1.064 \\
\textbf{ludcmp} & 8.992 & 1.060 & \textbf{statemate} & 8.293 & 1.063 \\
\textbf{matmult-float} & 18.41 & 0.953 & \textbf{stb\_perlin} & 17.88 & 0.951 \\
\textbf{matmult-int} & 8.141 & 1.067 & \textbf{stringsearch1} & 8.307 & 1.066 \\
\textbf{mergesort} & 8.244 & 1.080 & \textbf{strstr} & 7.332 & 1.073 \\
\textbf{miniz} & 21.14 & 1.033 & \textbf{tarai} & 7.414 & 1.074 \\
\textbf{minver} & 8.176 & 1.064 & \textbf{trio-snprintf} & 11.63 & 1.090 \\
\textbf{mont64} & 9.588 & 1.054 & \textbf{trio-sscanf} & 12.09 & 1.070 \\
\textbf{nbody} & 7.926 & 1.065 & \textbf{ud} & 8.088 & 1.067 \\
\textbf{ndes} & 8.846 & 1.060 & \textbf{whetstone} & 14.90 & 1.043 \\
\textbf{nettle-arcfour} & 7.738 & 1.066 & \textbf{wikisort} & 8.688 & 1.375 \\
\hline
\end{tabular}
\label{table:all-sizes}
\end{table}
\setlength{\tabcolsep}{6pt}

\end{arXivOnly}

\end{document}